\documentclass{PoS}
\usepackage{color}
\usepackage{amsmath}
\usepackage{graphicx}
\usepackage{amssymb}
\usepackage{bm}
\graphicspath{{./Figs/}}
\newcommand{\Comments}[1]{}
\title{Preweighting method in Monte-Carlo sampling with complex action
\\
--- Strong-Coupling Lattice QCD with $1/g^2$ corrections, as an example ---
\footnote{Preprint: YITP-15-96, KUNS-2595}}

\ShortTitle{Preweighting method in MC sampling with complex action}

\author{\speaker{Akira Ohnishi}
\\
	Yukawa Institute for Theoretical Physics, Kyoto University,
	Kyoto 606-8502, Japan\\
        E-mail: \email{ohnishi@yukawa.kyoto-u.ac.jp}}

\author{Terukazu Ichihara\\
	Department of Physics, Kyoto University,
	Kyoto 606-8502, Japan\\
	Yukawa Institute for Theoretical Physics, Kyoto University,
	Kyoto 606-8502, Japan\\
        E-mail: \email{t-ichi@ruby.scphys.kyoto-u.ac.jp}}

\abstract{
We investigate the QCD phase diagram
in the strong-coupling lattice QCD with fluctuation and $1/g^2$ effects
by using the auxiliary field Monte-Carlo simulations.
The complex phase of the Fermion determinant at finite chemical potential
is found to be suppressed by introducing a complex shift of integral path
for one of the auxiliary fields,
which corresponds to introducing a repulsive vector mean field for quarks.
The obtained phase diagram in the chiral limit
shows suppressed $T_c$ in the second order phase transition region
compared with the strong-coupling limit results.
We also argue that we can approximately guess the statistical weight
cancellation from the complex phase in advance in the case
where the complex phase distribution is Gaussian.
We demonstrate that correct expectation values are obtained
by using this guess in the importance sampling (preweighting).
}

\FullConference{The 33rd International Symposium on Lattice Field Theory\\
		14 -18 July 2015\\
		Kobe International Conference Center, Kobe, Japan*}

\begin{document}

%\section{...}
\section{Introduction}

QCD phase diagram is closely related to compact star phenomena
and can be accessed using heavy-ion collisions.
It is desirable to understand the QCD phase diagram
in non-perturbative first-principles theory, 
{\em i.e.} lattice QCD Monte-Carlo (LQCD-MC) simulations,
but LQCD-MC suffers from the sign problem
at finite chemical potential ($\mu$).
There are many methods proposed so far to circumvent the sign problem,
%Taylor expansion, 
%re-weighting, 
%analytic continuation from imaginary chemical potential,
%canonical method,
%fugacity expansion,
%histogram method,
%complex Langevin simulations,
%Lefschetz thimble,
%strong coupling,
but it is still difficult to attack low temperature ($T$) QCD matter.

The strong coupling lattice QCD (SC-LQCD) is 
one of the methods,
with which the phase boundary has been drawn 
using QCD~\cite{SCL-MF-PD,SC-LQCD-MF-PD,MDP-PD,AFMC,MDP-plaq,DS}.
In SC-LQCD, (spatial) link variables are integrated out first,
and we obtain the effective action consisting of color singlet composites
to a given order of $1/g^2$.
Then fluctuations of the complex phase from the link fluctuations
could be suppressed, namely the sign problem becomes milder.
For example, there is no sign problem 
in the mean field treatment~\cite{SCL-MF-PD,SC-LQCD-MF-PD},
and the phase diagram has been obtained
in the strong coupling limit ($\mathcal{O}(1/g^0)$)~\cite{SCL-MF-PD}
as well as with finite coupling effects~\cite{SC-LQCD-MF-PD}.
The sign problem appears when we include fluctuation effects,
but it is not severe in the strong coupling limit.
The phase boundary in the strong coupling limit with fluctuation effects
has been obtained in two independent method, 
the Monomer-Dimer-Polymer (MDP) simulations~\cite{MDP-PD}
and the Auxiliary Field Monte-Carlo (AFMC) method~\cite{AFMC},
and obtained phase boundaries agree with each other.
Towards the actual QCD phase diagram,
we need to take account of both finite coupling and fluctuation effects.
Recently, the plaquette effects on the phase boundary were evaluated
using the linear and exponential extrapolation 
in the MDP simulation~\cite{MDP-plaq}.
In AFMC, an effective action including $1/g^2$ contributions
was considered, but the sign problem was found to be severe~\cite{Lat2014}.
Thus the direct sampling method including the finite coupling corrections
should be further developed.

Another interesting technique to avoid the sign problem has been developed 
in the density of states method~\cite{DoS,Ejiri,Greensite}.
If the complex phase $\theta$ follows a Gaussian distribution,
%This Gaussian assumption would be plausible 
%provided that the complex phase appears randomly,
as examined %in LQCD-MC simulation 
in the heavy quark mass region~\cite{Ejiri},
%If this is the case,
we can integrate out $\theta$ and obtain the partition function
without the sign problem.
%One of the demerits of this method is that 
%we need to sample those configurations with very small statistical weight
%after integrating over the complex phase.
It is desirable to examine the $\theta$ distribution at small quark masses.
Furthermore,
it is preferable to judge the statistical weight in advance
in order to enhance numerical efficiency.

In this proceedings,
we discuss the QCD phase transition in SC-LQCD
with fluctuation and $1/g^2$ effects by using the AFMC method.
We introduce two ideas to suppress the complex phase spread;
the complex shift of auxiliary field integration path
and the {\em preweighting}.

\section{Auxiliary Field Monte-Carlo method in SC-LQCD
with $1/g^2$ corrections}
\label{sec:AFMC}

%\subsection{Effective action of auxiliary fields}
%\label{subsec:Seff}
We here consider the lattice QCD 
with one species of unrooted staggered fermion
%for color $\mathit{SU}(N_c)$ 
in the anisotropic Euclidean spacetime.
Throughout this paper, we work in the lattice unit $a=1$,
where $a$ is the spatial lattice spacing, 
and the case of color $\mathrm{SU}(N_c=3)$
in 3+1 dimension $(d=3)$ spacetime.
Temporal and spatial lattice sizes are denoted as $N_\tau$ and $L$,
respectively.

Starting from the lattice QCD partition function,
we obtain the effective action 
including the leading order 
(strong coupling limit (SCL), $\mathcal{O}(1/g^0)$) terms 
and next-to-leading order (NLO, $1/g^2$) corrections
by integrating out spatial link 
%using QCD~\cite{SCL-MF,SC-LQCD-MF,MDP-PD,AFMC,DS}.
variables~\cite{SCL-MF-PD,SC-LQCD-MF-PD,AFMC,SCL,Faldt,Jolicoeur},
\begin{align}
S_\mathrm{eff}
=&\frac12 \sum_x \left[ V^{+}_x(\mu) - V^{-}_x(\mu) \right]
- \frac{1}{4N_c\gamma^2} \sum_{x,j} M_x M_{x+\hat{j}}
+\frac{m_0}{\gamma} \sum_{x} M_x
\ ,\nonumber\\
+&\frac{\beta_\tau}{2} \sum_{x,j} \left[
	 V_x^+(\mu)\,V^-_{x+\hat{j}}(\mu)
	+V_{x+\hat{j}}^+(\mu)\,V^-_{x}(\mu)
	\right]
-\frac{\beta_s}{\gamma^4} \sum_{x,k,j,k\not=j}
	Q^{(j)}_x
	Q^{(j)}_{x+\hat{k}}
%M_x M_{x+\hat{j}} M_{x+\hat{k}} M_{x+\hat{j}+\hat{k}}
\label{Eq:Seff}
\ ,\\
V^{+}_x(\mu)=&  \bar{\chi}_x e^{\mu/\gamma^2}U_{0,x} \chi_{x+\hat{0}}
\ ,\ 
V^{-}_x(\mu)= \bar{\chi}_{x+\hat{0}} e^{-\mu/\gamma^2} U^\dagger_{0,x} \chi_x
\ ,\ 
M_x=\bar{\chi}_x \chi_x
\ ,\ 
Q^{(j)}_x=M_x M_{x+\hat{j}}
% M_{x+\hat{k}} M_{x+\hat{j}+\hat{k}}
\ ,
\end{align}
where
$\chi_x$ and $U_{0,x}$
represent the quark field and the temporal link variable,
respectively,
$V^{\pm}_x$, $M_x$ and $Q^{(j)}_x$ are mesonic composites,
%%%%%%%%%%%%
% (inverse square) gauge coupling constants appear in
%$\beta_\tau=1/2N_c^2g^2\gamma=\beta_g/4N_c^3\gamma$
$\beta_\tau=\beta_g/4N_c^3\gamma$,
and
%$\beta_s=1/16N_c^4g^2\gamma=\beta_g/32N_c^5\gamma$,
$\beta_s=\beta_g/32N_c^5\gamma$,
where $\beta_g=2N_c/g^2$.
%%%%%%%%%%%%
%Quark chemical potential $\mu$ is introduced
%in the form of the temporal component of vector potential.
%
We assume that the physical lattice spacing ratio 
is given as a function of the lattice anisotropy parameter $\gamma$ as
$f(\gamma)=a_\mathrm{s}^\mathrm{phys}/a_\tau^\mathrm{phys}=\gamma^2$,
%according to the discussion in the strong coupling limit~\cite{SCL-MF-PD},
then the temperature is given as $T=\gamma^2/N_\tau$~\cite{SCL-MF-PD}.

We obtain the effective action of quarks, temporal link variables,
and auxiliary fields in the bilinear form of quarks,
by applying the extended Hubbard-Stratonovich (EHS) transformation 
$\exp(\alpha AB) = \int d\phi d\phi^* \exp[-\alpha(\phi^*\phi+\phi^*A+\phi B)]$
several times,
%and obtain the effective action of quarks, temporal link variables,
%and auxiliary fields in the bilinear form of quarks,
\begin{align}
S_\mathrm{EHS}
=&\frac12 \sum_x \left(Z^-_x V_x^+ - Z^+_x V_x^-\right)
  +\frac{1}{\gamma} \sum_x m_x M_x
  +S_\mathrm{AF}
  \ ,\label{Eq:S_EHS}\\
m_x
=& m_0 + \frac{1}{4N_c} \sum_j (\sigma+i\varepsilon\pi)_{x\pm\hat{j}}
  +\beta_s \sum_j \left\{
   \varphi^{(j)*}_x (\Theta_x^{(j)})^{1/2}
  +\varphi^{(j)*}_{x-\hat{j}} (\Theta_{x-\hat{j}}^{(j)})^{1/2}
  \right\}
\ ,\nonumber\\
Z^-_x=& 1+\beta_\tau\sum_j(\omega-\varepsilon\Omega)_{x\pm\hat{j}}^*
\ ,\quad
Z^+_x= 1+\beta_\tau\sum_j(\omega+\varepsilon\Omega)_{x\pm\hat{j}}
\ ,\quad
\Theta^{(j)}_x=
  \sum_{k,k\not=j}
  (\sigma^{(j)}+i\varepsilon\pi^{(j)})_{x\pm \hat{k}}
\ ,\nonumber\\
S_\mathrm{AF}
=&\frac{L^3}{4N_c}
  \sum_{\bold{k},\tau,f(\bold{k})>0} f(\bold{k})
  \left[
  \sigma^*_{\bold{k},\tau} \sigma_{\bold{k},\tau}
  +\pi^*_{\bold{k},\tau} \pi_{\bold{k},\tau}
  \right]
  +\beta_\tau L^3 \sum_{\bold{k},\tau,f(\bold{k})>0}
  f(\bold{k})
  \left[
   \Omega_{\bold{k},\tau}^* \Omega_{\bold{k},\tau}
  +\omega_{\bold{k},\tau}^* \omega_{\bold{k},\tau}
  \right]
\nonumber\\
&+\beta_s L^3 \sum_{\bold{k},\tau,f^{(j)}(\bold{k})>0} f^{(j)}(\bold{k})
  \left[
  \sigma^{(j)*}_{\bold{k},\tau} \sigma^{(j)}_{\bold{k},\tau}
  +\pi^{(j)*}_{\bold{k},\tau} \pi^{(j)}_{\bold{k},\tau}
  \right]
  +\beta_s \sum_{x,j} \varphi^{(j)*}_x \varphi^{(j)}_x
\ ,\nonumber
\end{align}
where
$f(\bold{k})=\sum_j \cos k_j$
and
$f^{(j)}(\bold{k})=\sum_{k,k\not=j} \cos k_k$.
%and
%$x\pm\hat{j}$ implies that the sum for both $\pm$ directions to be taken.
%
SCL fields $(\sigma,\pi)$,
spatial $(\sigma^{(j)},\pi^{(j)})$ and temporal $(\Omega, \omega)$ NLO fields
are introduced to bosonize interaction terms,
$M_xM_{x+\hat{j}}$,
$Q^{(j)}_xQ^{(j)}_{x+\hat{k}}$,
and
$V^+_xV^-_{x+\hat{j}}+V^+_{x+\hat{j}}V^-_{x}$
terms in Eq.~\eqref{Eq:Seff}, respectively.
The other spatial NLO field $\varphi^{(j)}$ is introduced
in the second step bosonization of the eight-Fermi term,
$Q^{(j)}_xQ^{(j)}_{x+\hat{k}}$.
It should be noted that
%we bosonize terms 
%with $(\bold{k},-\bold{k})$ and $(-\bold{k},\bold{k})$ simultaneously,
%then the conditions
$\sigma$, $\pi$, $\sigma^{(j)}$ and $\pi^{(j)}$ fields
are real in the coordinate representation.
%$\sigma_{-\bold{k},\tau}=\sigma^*_{\bold{k},\tau}$
%and $\pi_{-\bold{k},\tau}=\pi^*_{\bold{k},\tau}$
%are satisfied automatically.
%The same property applies to $\sigma^{(j)}$ and $\pi^{(j)}$,
%and we find the coordinate representation of these fields become real, 
%$\sigma_x, \pi_x, \sigma^{(j)}_x, \pi^{(j)}_x \in \mathbb{R}$.

%In the above action, the fermion matrix is
%block diagonal for each spatial point,
%decomposed into that for each spatial point,
%and we can calculate the fermion determinant
%by using the recursion technique~\cite{Faldt}.
We can integrate out quarks and temporal links analytically,
and the partition function and the effective action in AFMC
are found to be,
\begin{align}
\mathcal{Z}_\mathrm{AFMC}
=& \int \mathcal{D}\Phi
\exp\left(-S_\mathrm{AFMC}[\Phi]\right)
\quad(
\Phi=\left\{
\sigma,\pi,\sigma^{(j)},\pi^{(j)},\varphi,\varphi^*,\Omega,\Omega^*,
\omega,\omega^*
\right\})
\ ,\\
S_\mathrm{AFMC}
=& S_\mathrm{AF} - \sum_{\bold{x}} \log\left[R(\bold{x})\right]
\ ,\label{Eq:S_AFMC}\quad
R(\bold{x})=
Z^3\left[
(X_{N_\tau}/Z)^3
-2(X_{N_\tau}/Z)
+2 \cosh(N_c\widetilde{\mu}/T)
\right]
%\ ,\\
%Z=& \left[
%\prod_\tau Z_{\bold{x},\tau}^- Z_{\bold{x},\tau}^+
%\right]^{1/2}
%\ ,\quad
%\widetilde{\mu}/T=\mu/T
%+\frac12\sum_\tau\log\left(Z_{\bold{x},\tau}^-/Z_{\bold{x},\tau}^+\right)
%\ .
\ ,
\end{align}
where 
$Z= [\prod_\tau Z_{\bold{x},\tau}^- Z_{\bold{x},\tau}^+]^{1/2}$
and 
$\widetilde{\mu}/T=\mu/T+\frac12\sum_\tau\log\left(Z_{\bold{x},\tau}^-/Z_{\bold{x},\tau}^+\right)$.
$X_{N_\tau}$ is obtained by using the recursion formula,
$X_N(1;N) = B_N(1;N)+\gamma_N B_{N-2}(2;N-1)$, 
$B_N(1;N) = I_N B_{N-1}(1;N-1) + \gamma_{N-1} B_{N-2}(1;N-2) \ (N>3)$,
$B_1(1;1) = I_1$, 
and 
$B_2(1;2) = I_1 I_2 + \gamma_1$,
where
$I_\tau= 2m_{\bold{x},\tau}/\gamma$ and 
$\gamma_\tau= Z^-_{\bold{x},\tau} Z^+_{\bold{x},\tau}$~\cite{Faldt}.

%It should be noted that there is no approximation
%after introducing the NLO effective action in QCD, Eq.~\eqref{Eq:Seff}.

%\subsection{Phase transition with $1/g^2$ corrections}
%\label{subsec:results}

\begin{figure}
\begin{center}
\includegraphics[width=7cm]{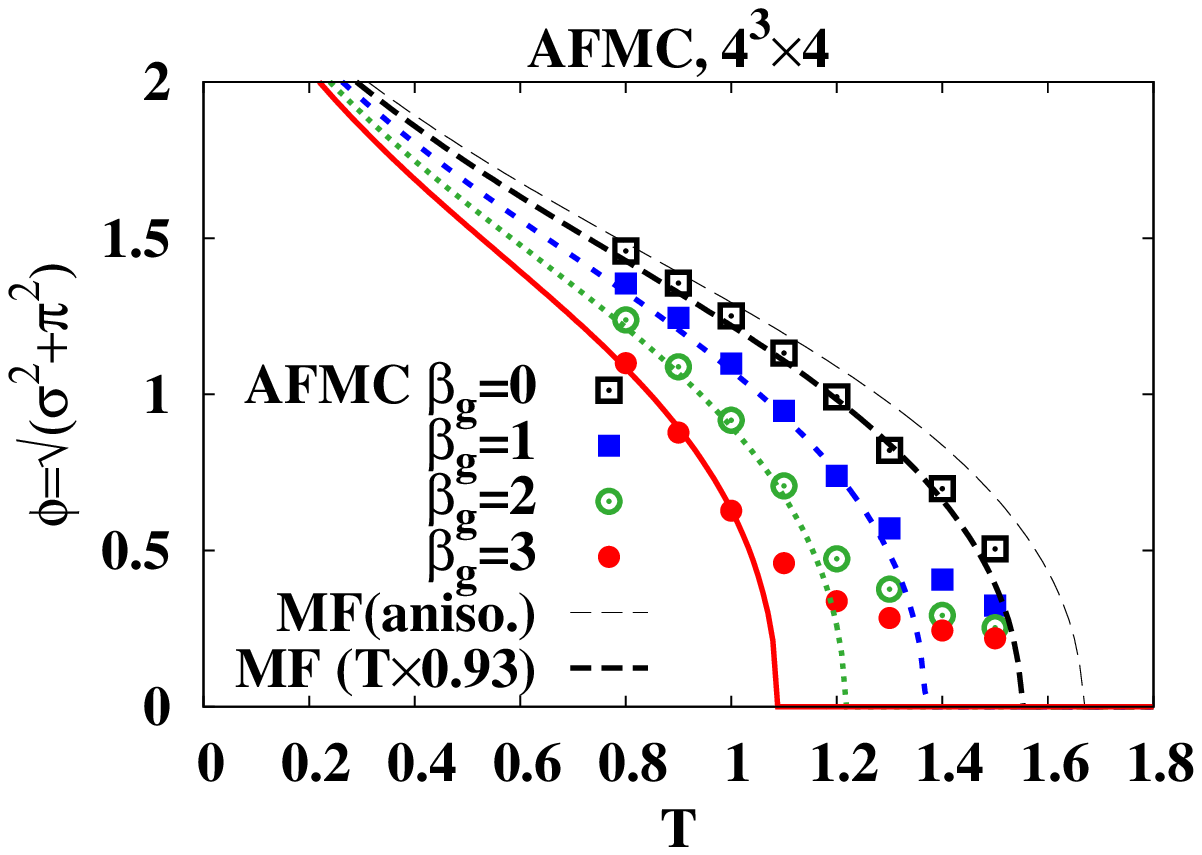}%
\includegraphics[width=7cm]{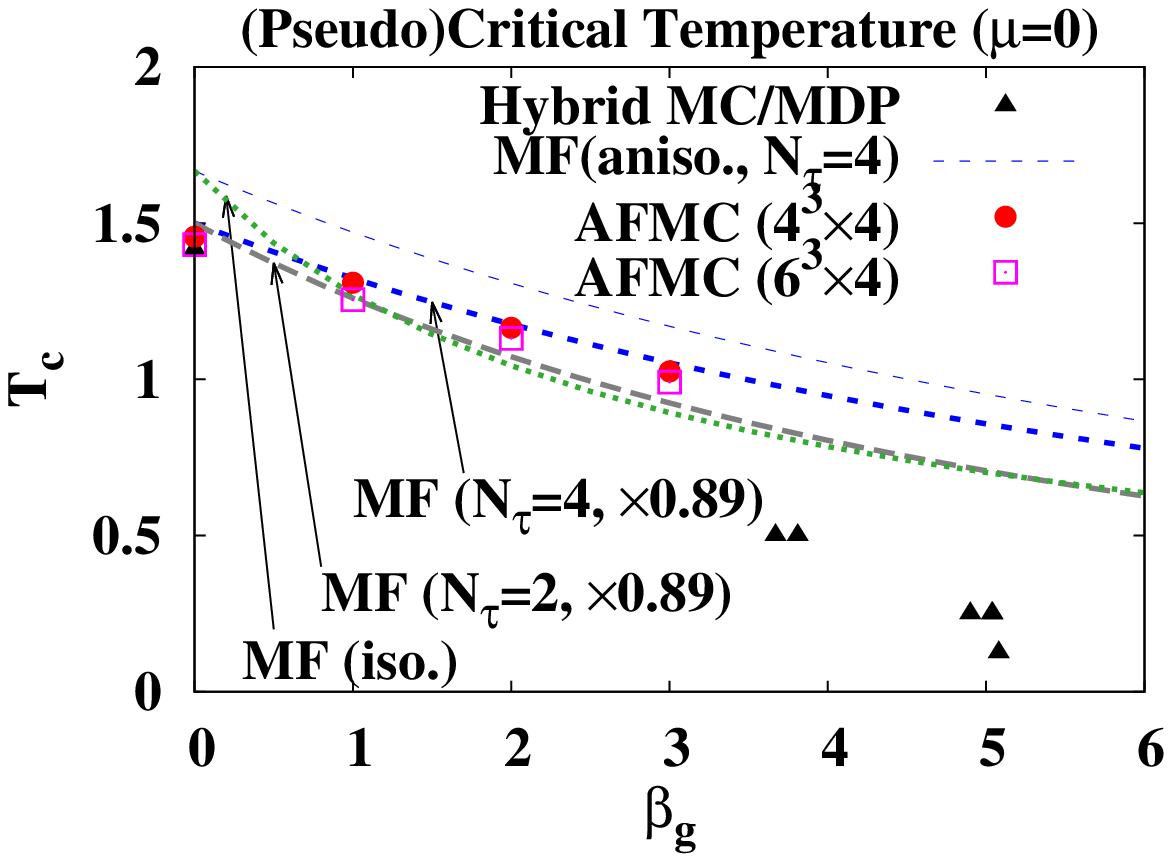}
\end{center}
\caption{Chiral condensate (left) and critical temperature (right)
at $\mu=0$.}
\label{Fig:mu0}
\end{figure}

We now perform Monte-Carlo calculation using 
the auxiliary field effective action Eq.~\eqref{Eq:S_AFMC}.
We have made calculations at $\beta_g=0,1,2,3$
on $4^3\times4$ and $6^3\times4$ lattices
in the chiral limit ($m_0=0$).
In the left panel of Fig.~\ref{Fig:mu0},
we show the chiral condensate at $\mu=0$ as a function of $T$
on a $4^3\times4$ lattice at $\beta_g=0,1,2,3$
in comparison with the mean field results on anisotropic lattice
with $N_\tau=4$.
We define the chiral circle radius
%$\phi=\left\langle\sqrt{\sigma^2_{k=0}+\pi^2_{k=0}}\right\rangle$
$\phi=\sqrt{\sigma^2_{k=0}+\pi^2_{k=0}}$
as the chiral condensate, since we work in the chiral limit.
The chiral condensate is suppressed from the mean field results,
% at the same temperature,
and it roughly corresponds to the mean field result
at $T_\mathrm{MF}=T/0.93$.
In the right panel of Fig.~\ref{Fig:mu0},
we show the critical temperature $T_c$ as a function of $\beta_g$.
We have made a function fit to the chiral condensate in AFMC,
and the critical temperature is defined as the maximum of $-d\phi/dT$
of the fitted function.
The behavior of $T_c$ is again consistent with the scaled mean field
results on an anisotropic lattice,
$T_c(\mathrm{AFMC})\simeq 0.89 T_c(\mathrm{MF, aniso.}, N_\tau=4)$.
%This point seems to disagree with the MDP results,
%where $T_c$ decreases more rapidly as a function of $\beta_g$.

\begin{figure}
\begin{center}
\includegraphics[width=7cm]{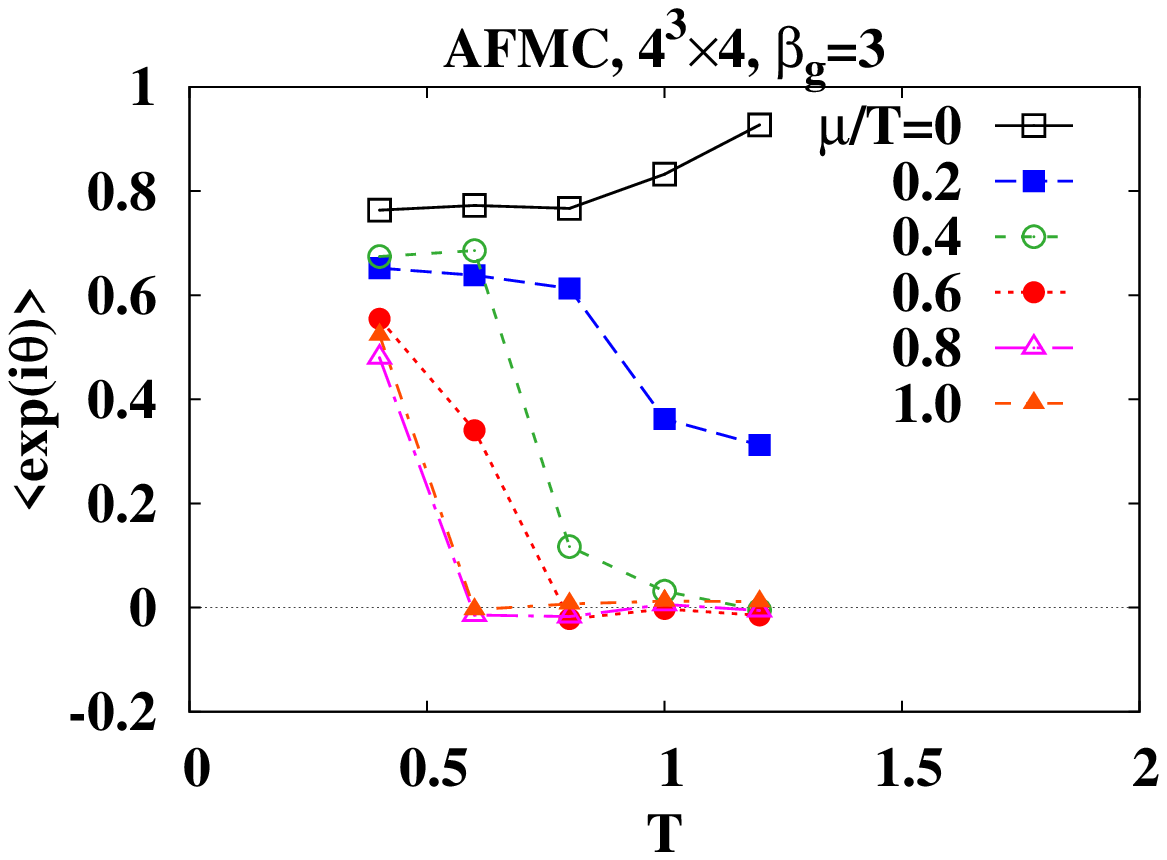}%
\includegraphics[width=7cm]{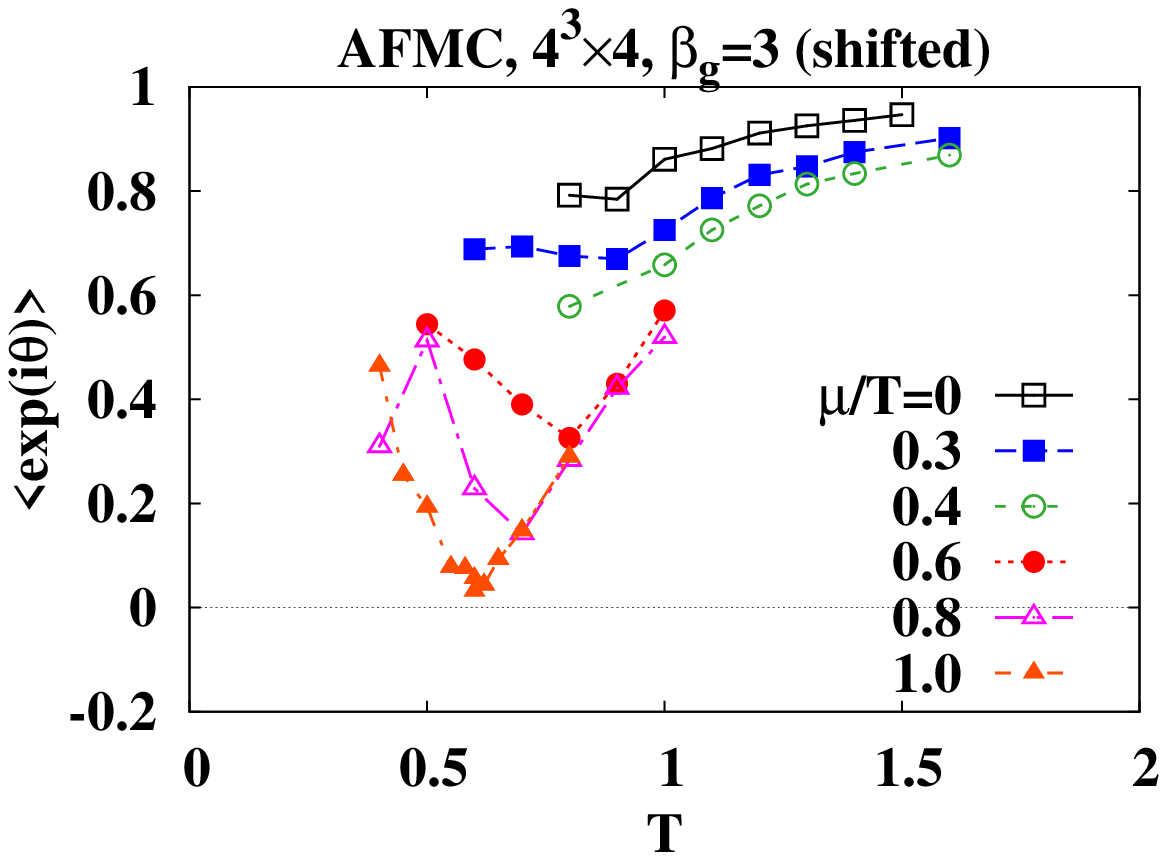}\\
\includegraphics[width=7cm]{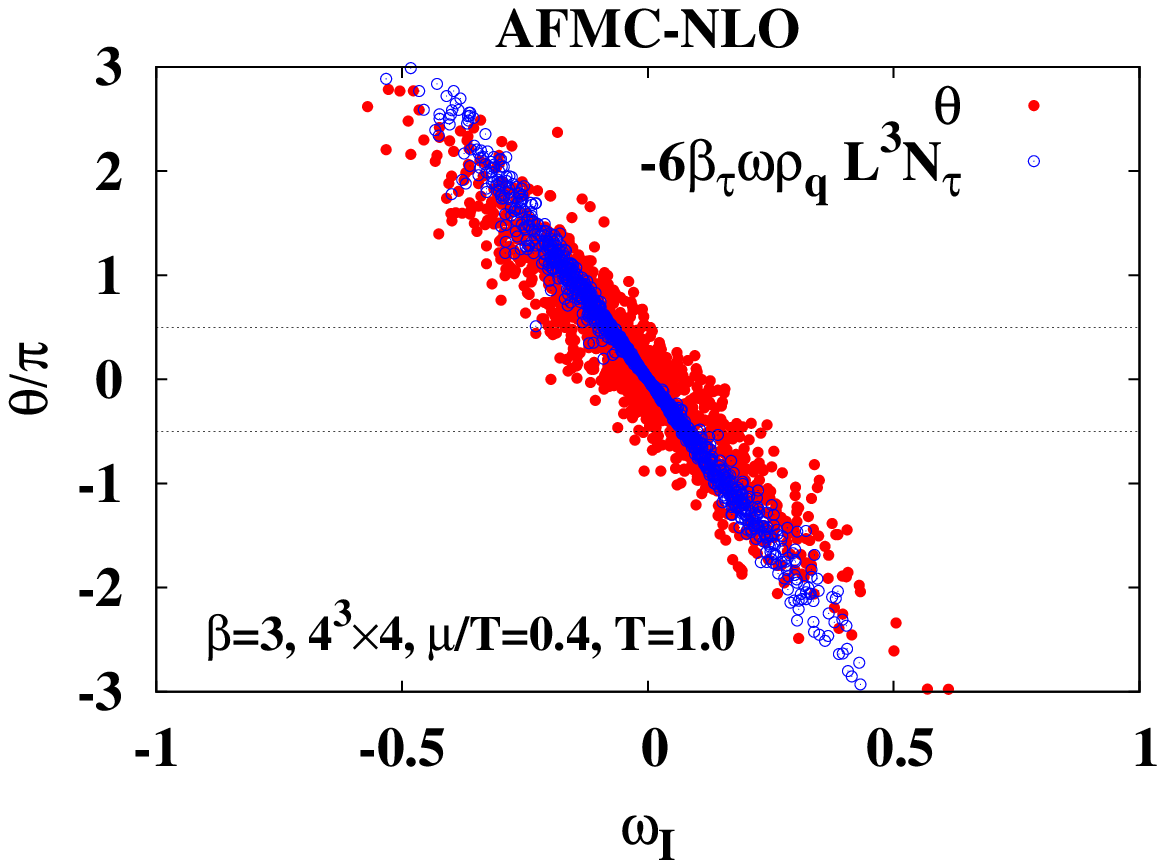}%
\includegraphics[width=7cm]{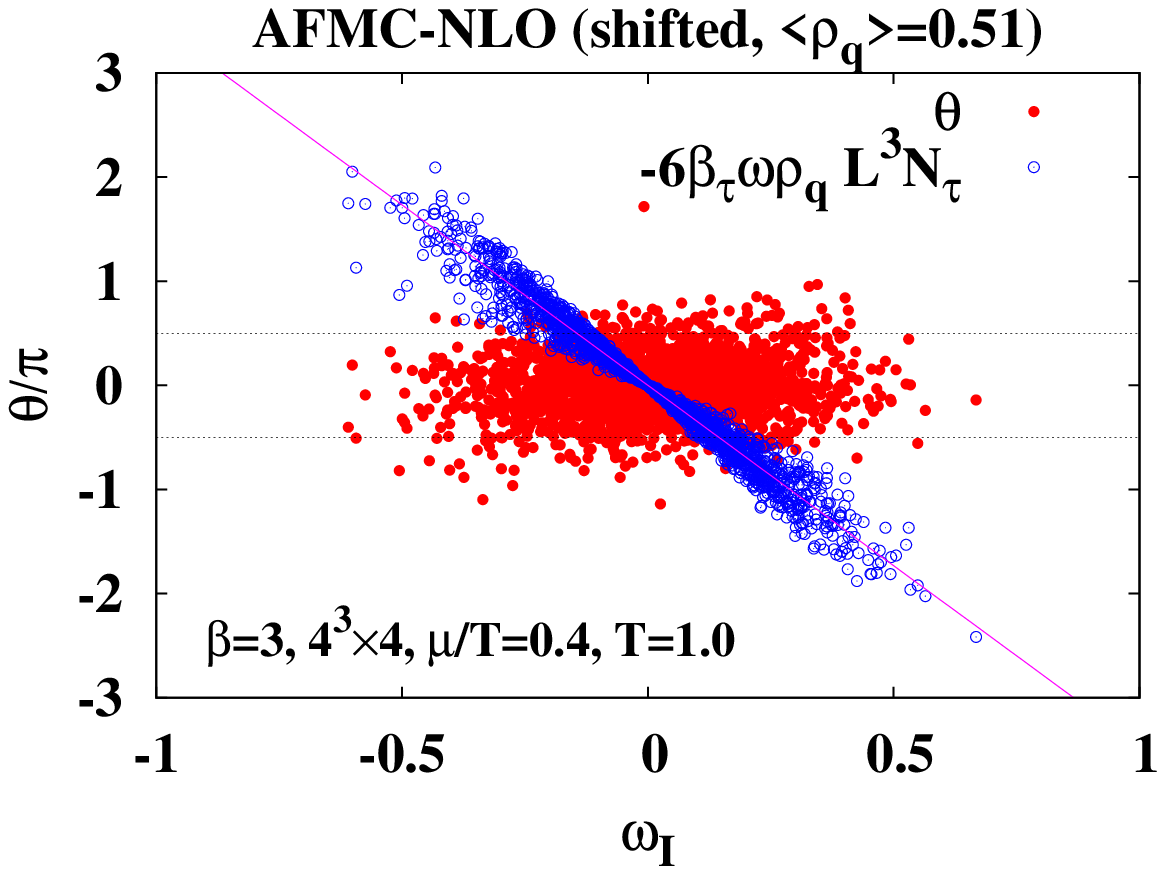}
\end{center}
\caption{Average phase factor (upper panels)
and correlation between $\omega_I$ and the complex phase (lower panels)
in the original (left) and subtracted (right) action.
}
\label{Fig:rotate}
\end{figure}

We have a sign problem in AFMC coming from the bosonization;
decomposing the product of different composites requires
introducing complex fields,
as seen in the position dependent mass $m_x$ in Eq.~\eqref{Eq:S_EHS},
then the fermion determinant becomes a complex number.
As shown in the upper left panel of Fig.~\ref{Fig:rotate},
the average phase factor at $\mu/T=0$ is large enough,
$\langle\exp(i\theta)\rangle>0.7$.
%in AFMC
%as a function of $T$ at $\mu/T=0, 0.2, 0.4, 0.6, 0.8$ and $1.0$
%in the upper left panel of
%At $\mu/T=0$, the average phase factor is large enough
%($\langle\exp(i\theta)\rangle>0.7$), and the sign problem is not severe.
%
At $\mu/T>0.4$, however,
it suddenly collapses at around $T\sim 0.7$.
%where the quark number density becomes finite.
%
This collapse comes from finite density.
%is related to the repulsion from vector potential.
Let us examine this point.
We focus our attention to those terms involving the imaginary part 
of $\omega$ at zero momentum in the EHS action Eq.~\eqref{Eq:S_EHS},
\begin{align}
S_{\omega_I} = \frac12 C \omega_I^2 + i C \omega_I \rho_q\ 
\quad
(C=6\beta_\tau L^3 N_\tau)\ ,
\end{align}
where $\omega_I=\sum_\tau \mathrm{Im} \omega_{\bold{k}=0,\tau} / N_\tau$,
and $\rho_q$ is the quark number density.
When $\rho_q$ is finite, the above term gives rise to a complex phase
of $\theta=-C\omega\rho_q$.
As shown in the lower left panel of Fig.~\ref{Fig:rotate},
%we show the correlation of $\omega_I$ and $\theta$
%together with $-C\omega_I\rho_q$.
%The agreement of these correlations tells us that
the above term dominates the complex phase.
Fortunately, this sign problem is a kind of textbook example;
%and the prescription is known
%--- 
By shifting the integral path to the imaginary direction
%Thus we shift 
$\omega_I \to \omega_I-i\rho_q$,
%and the subtracted action is found to be
we can remove the imaginary part
$S_{\omega_I} \to C \omega_I^2/2 + C \rho_q^2/2$.
We tune the shift constant $\langle\rho_q\rangle$ for each $(T,\mu)$ carefully,
then $\theta$ is suppressed
and $\langle\exp(i\theta)\rangle$ becomes larger 
%and the average phase factor
as shown in the right panels of Fig.~\ref{Fig:rotate}.

\begin{figure}
\begin{center}
\includegraphics[width=7cm]{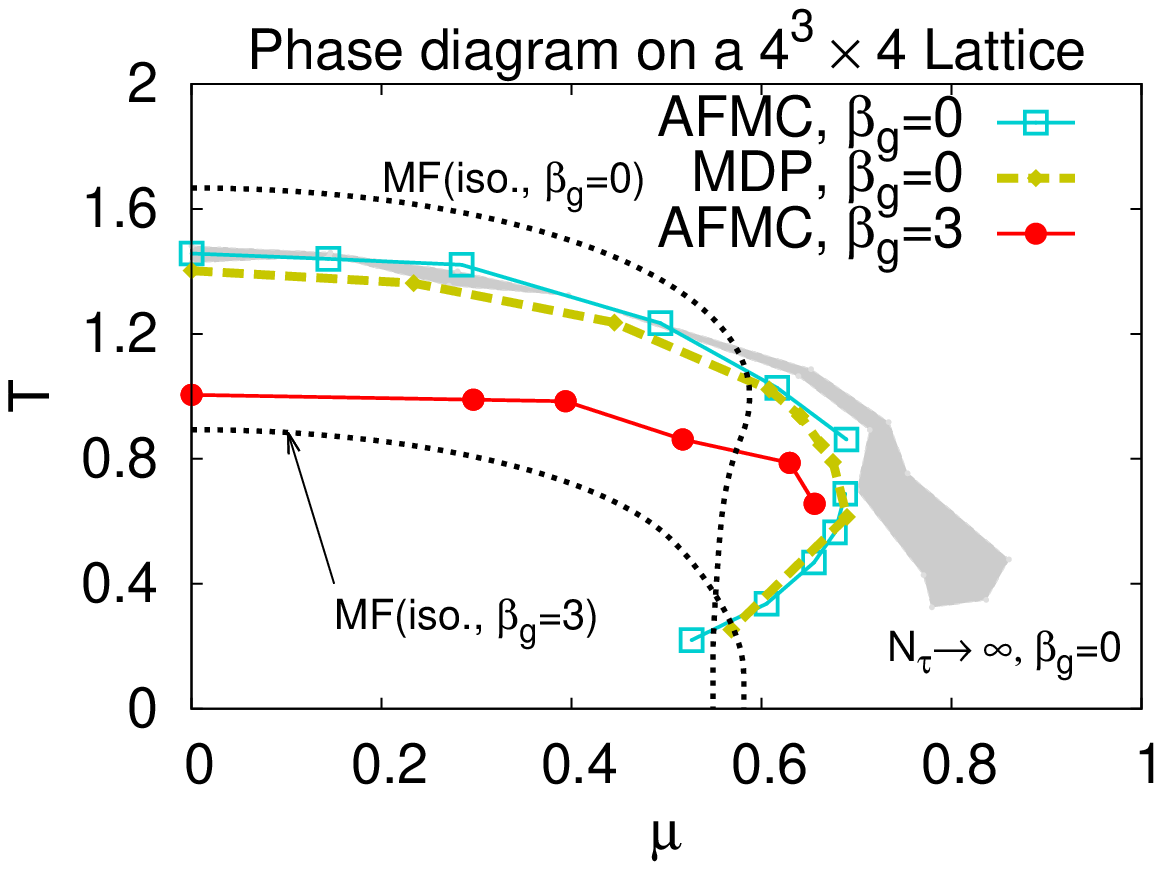}%
\includegraphics[width=7cm]{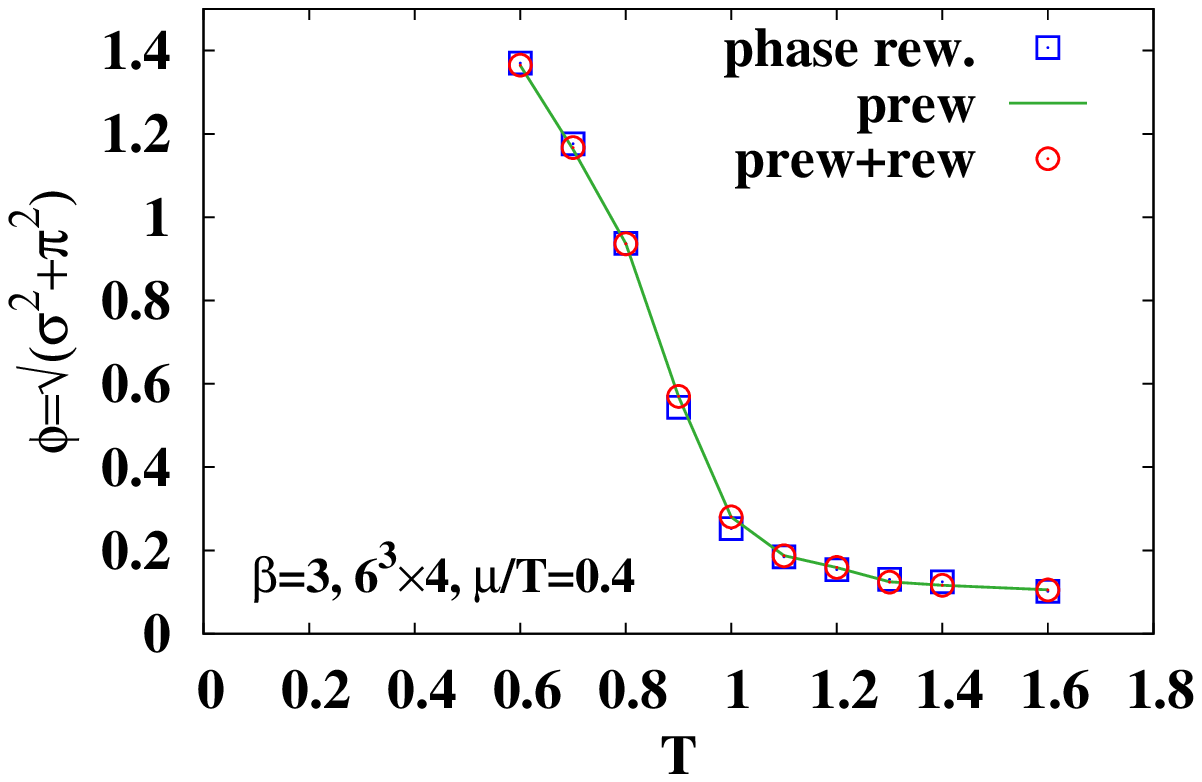}%
\end{center}
\caption{Phase diagram in SC-LQCD with $1/g^2$ correction at $\beta_g=3$
(left) and the chiral condensate in preweighting and reweighting (right).}
\label{Fig:PBprew}
\end{figure}

We show the phase boundary at $\beta_g=3$
in the left panel of Fig.~\ref{Fig:PBprew},
in comparison with the phase boundary in SCL.
The phase boundary at small $\mu/T$ seems to be flatter
compared with the SCL boundary.
This is consistent with the previous results~\cite{MDP-plaq,Lat2014}.
At $\mu/T>1$ where the first order phase transition would appear,
a constant shift is not enough.
When the high-density Wigner phase and the low-density Nambu-Goldstone phase
coexist, we need to introduce configuration dependent shifts.
%In the above shift, we have assumed that the density fluctuation 
%in different configurations is small.
%When the density fluctuation is large as in the coexisting region
%of the high-density Wigner phase and the low-density Nambu-Goldstone phase,
%two local minima appears in the density
%and the complex phase cannot be removed by a constant shift of $\omega_I$.
%It is important to introduce configuration dependent shifts.
Work in this direction is in progress.

\section{Preweighting}

For a calculation on larger lattices,
complex shift of the integration path of one auxiliary field is not enough.
%Even if we can suppress the complex phase by shifting the integration path,
%We still have remaining complex phases whose variance $\Delta^2$
%is finite and would be proportional to the system volume.
%Then we cannot avoid strong weight cancellation
One of the ideas to circumvent the sign problem is
to apply the density of states approach~\cite{DoS,Ejiri,Greensite},
where
%In the density of states approach,
we obtain the density of states for a given variable $x$,
% to take $x=x_i$,
and evaluate the average phase factor in each bin of $x$.
%at $x=x_i$.
Leading order truncation of the cumulant expansion in $\theta$
corresponds to assuming that the complex phase distribution
is Gaussian~\cite{Ejiri}.
%\bibitem{Greensite:2013ska}
If this is the case, we can analytically integrate out $\theta$
and avoid the sign problem.
Let us assume that the effective action is quadratic
in $\theta$,
$S[\Phi,\theta]=S_\Phi + \theta^2/2\Delta^2_\Phi-i\theta$,
then the partition function is calculated as
\begin{align}
\mathcal{Z}
=&\int \mathcal{D}\Phi d\theta\, e^{-S[\Phi,\theta]}
=\int \mathcal{D}\Phi\, 
\sqrt{2\pi}\Delta_\Phi\,
e^{-S_\Phi-\Delta^2_\Phi/2}
\ ,
\end{align}
where $\Phi$ shows variables other than $\theta$,
and $\Delta^2_\Phi$ is a variance of $\theta$ depending on $\Phi$.
In the heavy-quark mass case,
the complex phase $\theta$ was evaluated 
by using the Taylor expansion, % around $\mu/T=0$,
and the reweighted $\theta$ distribution was examined to be well approximated
by a Gaussian~\cite{Ejiri}.
In order to generate MC samples at finite $\mu$ directly,
we have an overlap problem and a problem of numerical cost.
In standard phase quenched lattice QCD simulations,
ignoring the complex phase leads to misidentification
of $U$ and $U^\dagger$ and of pions and diquarks.
The misidentification gives rise to the overlap problem;
the appearance of pion condensed phase
and superposed Lorentzian distribution of $\theta$~\cite{Lombardo}
at $\mu>m_\pi/2$ at low temperature,
instead of baryonic matter.
In SC-LQCD, we integrate out link variables analytically,
and the pion condensed phase does not appear.
By comparison, the cost problem remains.
Configurations of $\Phi$ with large $\Delta^2_\Phi$ are suppressed
by the factor $\exp(-\Delta^2_\Phi/2)$,
but we do not know $\Delta^2_\Phi$ in advance.

Let us now introduce a {\em preweighting} factor,
which suppresses configurations with large $\theta$,
\begin{align}
\mathcal{Z}_\mathrm{prew}
\equiv& \int \mathcal{D}\Phi\,d\theta\,
e^{-S_\Phi-\theta^2/2\Delta^2_\Phi}
e^{-f(\theta)}
= \int \mathcal{D}\Phi\,
e^{-S_\Phi}
F(\Delta_\Phi)
\ ,
\end{align}
where $F(\Delta)=\int d\theta\,e^{-\theta^2/2\Delta^2-f(\theta)}$.
If we find a function $f(\theta)$ which satisfies
$F(\Delta)/\sqrt{2\pi}\Delta=\exp(-\Delta^2/2)$,
we can obtain a correct partition function.
For small values of $\Delta$, the preweighting function
$f(\theta)=\theta^2/2+\theta^4/12+\theta^6/45+17\theta^8/1260+\mathcal{O}(\theta^{10})$,
is found to give a good approximation,
$F(\Delta)/\sqrt{2\pi}\Delta=\exp(-\Delta^2)+\mathcal{O}(\Delta^{10})$.
Since $F(\Delta)$ becomes a constant at large $\Delta$,
$F(\Delta)\to\int d\theta e^{-f}\ (\Delta \to \infty)$,
and the preweighting partition function deviates from the correct one,
we need to perform reweighting afterwards.
%Nevertheless it is practically efficient to introduce the preweighting function
We first obtain configurations in phase-quenched importance sampling
with the preweighting factor,
and calculate observables $\langle\hat{\phi}\rangle$
with the reweighting method,
$\langle \hat{\phi} \rangle
=\sum_n \phi_n R(\Delta_n) / \sum_n R(\Delta_n)$
with 
$R(\Delta)=\sqrt{2\pi}\Delta\,e^{-\Delta^2} / F(\Delta)$,
where $n$ denotes one configuration
and the variance $\Delta_n^2$ is obtained in a bin of $\phi$
to which $\phi_n$ belongs.
%prewsig-m04.eps

We have applied the preweighting + reweighting
to SC-LQCD with $1/g^2$ corrections.
We have confirmed that the complex phase distribution
is approximated by Gaussian.
In the right panel of Fig.~\ref{Fig:PBprew}
we show calculated results of chiral condensate $\phi$.
Since the complex shifted path is applied
and the complex phase variance is small in this case,
the results of the phase reweighting, preweighting,
and preweighting+reweighting agree with each other.
Comparison under more severe condition would be necessary
to verify the usefulness of the preweighting method.

\section{Summary}

In this proceedings, we have investigated the QCD phase diagram
by using the auxiliary field Monte-Carlo simulations
of the strong-coupling lattice QCD with $1/g^2$ corrections.
We have found that a complex shift of integral path
for one of the auxiliary fields ($\omega_I$) suppresses 
the complex phase of the Fermion determinant at finite $\mu$,
which corresponds to introducing a repulsive vector mean field for quarks.
The obtained phase diagram in the chiral limit
shows suppressed $T_c$ in the second order phase transition region
compared with the strong-coupling limit results.
as suggested in previous works using the mean field 
approximation~\cite{SC-LQCD-MF-PD}
and the reweighting method in the monomer-dimer-polymer 
simulations~\cite{MDP-plaq}.
We have also proposed a preweighting method.
%in which we can circumvent the sign problem.
When the distribution of the complex phase $\theta$ is Gaussian
and the variance of $\theta$ is small,
we can obtain a correct expectation value of an observable
from the phase quenched configurations obtained
by introducing the preweighting factor in the sampling process.
We need to combine preweighting and reweighting in the large variance cases.
%In the auxiliary field Monte-Carlo simulation
%of the strong-coupling lattice QCD with $1/g^2$ effects,
We have demonstrated that the preweighting+reweighting works well.

\bigskip

This work is supported in part by 
JSPS/MEXT Grant Numbers
 24540271, %((C) AO, T.Kunihiro, K.Morita),
 15H03663, %((B) PI:A.Nakamura),
 15K05079, %((C) AO, Y.Nara)),
 24105001, %(NSmatter X01)
and
 24105008. %(NSmatter D01)
%and by the Yukawa International Program for Quark-Hadron Sciences.
TI is supported by the Grants-in-Aid for JSPS Fellows (No.25-2059).

\end{document}